\UseRawInputEncoding 
\documentclass[12pt]{spieman}  
\usepackage{amsmath,amsfonts,amssymb}
\usepackage{graphicx}
\usepackage{setspace}
\usepackage{tocloft}
\usepackage{array}
\usepackage{xcolor}

\title{Instrument Signature Removal and Calibration Products for the Rubin Legacy Survey of Space and Time}

\author[a,b,*]{Andr\'es A. Plazas Malag\'on}
\author[d]{Chris Waters}
\author[c]{Alex Broughton}
\author[a, b]{Eli Rykoff}
\author[a,b]{Agn\`es Fert\'e}
\author[e]{Merlin Fisher-Levine}
\author[d]{Robert Lupton}
\affil[a]{SLAC National Accelerator Laboratory, Menlo Park, CA, USA}
\affil[b]{Kavli Institute for Particle Astrophysics and Cosmology, Stanford University, Stanford, CA, USA}
\affil[c]{Department of Physics and Astronomy, University of California-Irvine, Irvine, California, USA 92617}
\affil[d]{Department of Astrophysical Sciences, Princeton University, Princeton, NJ, USA}
\affil[e]{Rubin Observatory Project Office, 950 N. Cherry Ave., Tucson, AZ, USA}

\cftpagenumbersoff{figure}
\cftpagenumbersoff{table} 
\begin{document} 
\maketitle

\begin{abstract}
The Vera C. Rubin Legacy Survey of Space and Time (LSST) will conduct an unprecedented optical survey of the southern sky, imaging the entire available sky every few nights for 10 years. To achieve its ambitious science goals of probing dark energy and dark matter, mapping the Milky Way, and exploring the transient optical sky, {the systematic errors in the LSST data must be exquisitely controlled}. Instrument signature removal (ISR) is a critical early step in LSST data processing to remove inherent camera effects from the raw images and produce accurate representations of the incoming light. This paper describes the current state of the ISR pipelines implemented in the LSST Science Pipelines software. The key steps in ISR are outlined and the process of generating and verifying the necessary calibration products to carry out ISR is also discussed. Finally, an overview is given of how the Rubin data management system utilizes a data Butler and calibration collections to organize datasets and match images to appropriate calibrations during processing. Precise ISR will be essential to realize {the potential of LSST to revolutionize astrophysics}.\end{abstract}


{\noindent \footnotesize{*}Andr\'es A. Plazas Malag\'on,  \linkable{plazas@slac.stanford.edu} }


\section{Introduction}
\label{sect:intro}  
The Vera C. Rubin Observatory, currently under construction on Cerro Pachón in Chile, will undertake the Legacy Survey of Space and Time (LSST) using the state-of-the-art Simonyi Survey Telescope. With an 8.4-meter primary mirror and a 3.2-gigapixel camera (LSSTCam), the LSST aims to conduct an unprecedented survey of over 18,000 square degrees of the southern sky using six different filters spanning the optical and near-infrared spectrum. The LSST has four main science goals: probing dark energy and dark matter, creating an inventory of the solar system, exploring the transient optical sky, and {inferring} the structure and evolution of the Milky Way galaxy\cite{Ivezi__2019}. To achieve these goals, the LSST area will be observed more than 800 \textbf{\textcolor{black}{times}} in 10 years.  

The focal plane of the LSSTCam consists of 189 thick (100 microns), fully-depleted, back-illuminated science Charge-Coupled Devices (CCDs) \cite{holland03,holland09,holland14}, in addition to 8 CCDs for guiding and 4 split CCDs for wavefront measurements \cite{O'Connor16,Arndt10}. The science CCDs are arranged in 21 rafts, {each powering and controlling nine CCDs}. Each CCD is {divided into sixteen segments, each with a separate readout register and amplifier,} for low-noise, fast parallel readout, resulting in 3024 image segments for the science detectors. The LSSTCam focal plane contains CCDs fabricated by two vendors, the ITL STA3800C from the University of Arizona Imaging Technology Laboratory (ITL\cite{lesser17,itl_website})  and the E2V CCD250 from Teledyne E2V.\cite{e2v_website}

To attain the precision necessary for the LSST's ambitious science goals, the systematic errors in the data must be exquisitely controlled.  {{\textcolor{black}{Survey requirements are typically defined in terms of key observables, including photometry, astrometry, and the measurement of the point-spread function (PSF) and galaxy shapes. For astrometry, single-image positional uncertainties are expected to be 10 milliarcseconds (mas) (0.05 pixels), which ensures proper motion precision of 0.2 mas yr$^{-1}$ and parallax accuracy of 1.0 mas over the ten-year duration of the LSST \cite{lsst2011}. Photometric accuracy must be better than 10 millimagnitudes (mmag) to meet the survey's scientific goals \cite{Ivezi__2019}. PSF shape uncertainties in the final galaxy shear catalog are designed to be dominated by statistical errors, characterized by a dimensionless ellipticity scatter of $\sim 10^{-3}$, setting an upper limit on systematic error contributions. Similarly, the PSF fractional size bias must not exceed $10^{-3}$ across the full duration of the survey to satisfy cosmic shear requirements \cite{desc2018}}}}

The raw data from the camera contain inherent instrument signatures caused by effects such as bias levels, dark current, and amplifier noise. {\textcolor{black}{Instrument signature removal (ISR) is a critical initial step in the processing of LSST data products---comprising \textit{Prompt Products} (released within 24 hours for rapid follow-up of time-domain events), \textit{Data Release Products} (annual, including calibrated images, deep coadds, and catalogs for static sky and time-domain science), and \textit{User-Generated Products} \cite{lsstDPDP}---designed to eliminate these camera-induced effects and produce accurate representations of the incoming light. This process uses specific calibration images and corrections to remove instrumental artifacts from the raw data.}} {\textcolor{black}{For distribution of alerts to brokers, the entire Prompt Products processing pipeline---including ISR, image differencing, source detection, and alert distribution---must be completed within 60 seconds of image acquisition, ensuring minimal overhead. This stringent requirement is crucial for enabling the rapid follow-up of transient and time-critical astronomical events, such as supernovae, gamma-ray bursts, and near-Earth asteroids. Performance tests using data from the most recent commissioning campaign with LSSTComCam\cite{comcam_commissioning,howard2018} (a commissioning camera consisting of 9 ITL CCDs) showed a median runtime of approximately 14 seconds for the ISR component of the Prompt Products pipeline (K. Findeisen [U. of Washington / Rubin], private communication)}. }

{\textcolor{black}{The construction of calibrations for ISR is verified through a series of tests and metrics that assess the quality of the calibrations. Beyond ISR, additional tests can be conducted to address residual trends not fully removed by ISR (e.g., long-term trends), depending on the requirements of each science case. For example, cosmic shear analyses are often accompanied by PSF modeling null tests (e.g., correlations between size and ellipticity residuals) and may exclude the bright end of the stellar sample to mitigate systematic instrumental effects such as the brighter-fatter effect.\cite{schutt2025,li2022,jarvis2021} All LSST data (raw and calibrated images, yearly data releases) will be archived in their entirety, and improved correction algorithms can be incorporated into the LSST Science Pipelines for annual data reprocessing. \cite{lsstDPDP}}}

This paper is structured as follows: Section \ref{sec:isr} describes the steps in {ISR} currently implemented in the LSST Science Pipelines \cite{bosch2018, bosch2019} codebase, including {steps planned for future development
}. Section \ref{sec:calib} describes the generation, construction, and verification process of the necessary calibrations for ISR in the LSST. Section \ref{sec:butler} concludes {with} a general description of the Rubin data management paradigm and how the pipelines are constructed and executed.

    \begin{figure}
      \centering
        \includegraphics[width=\linewidth]{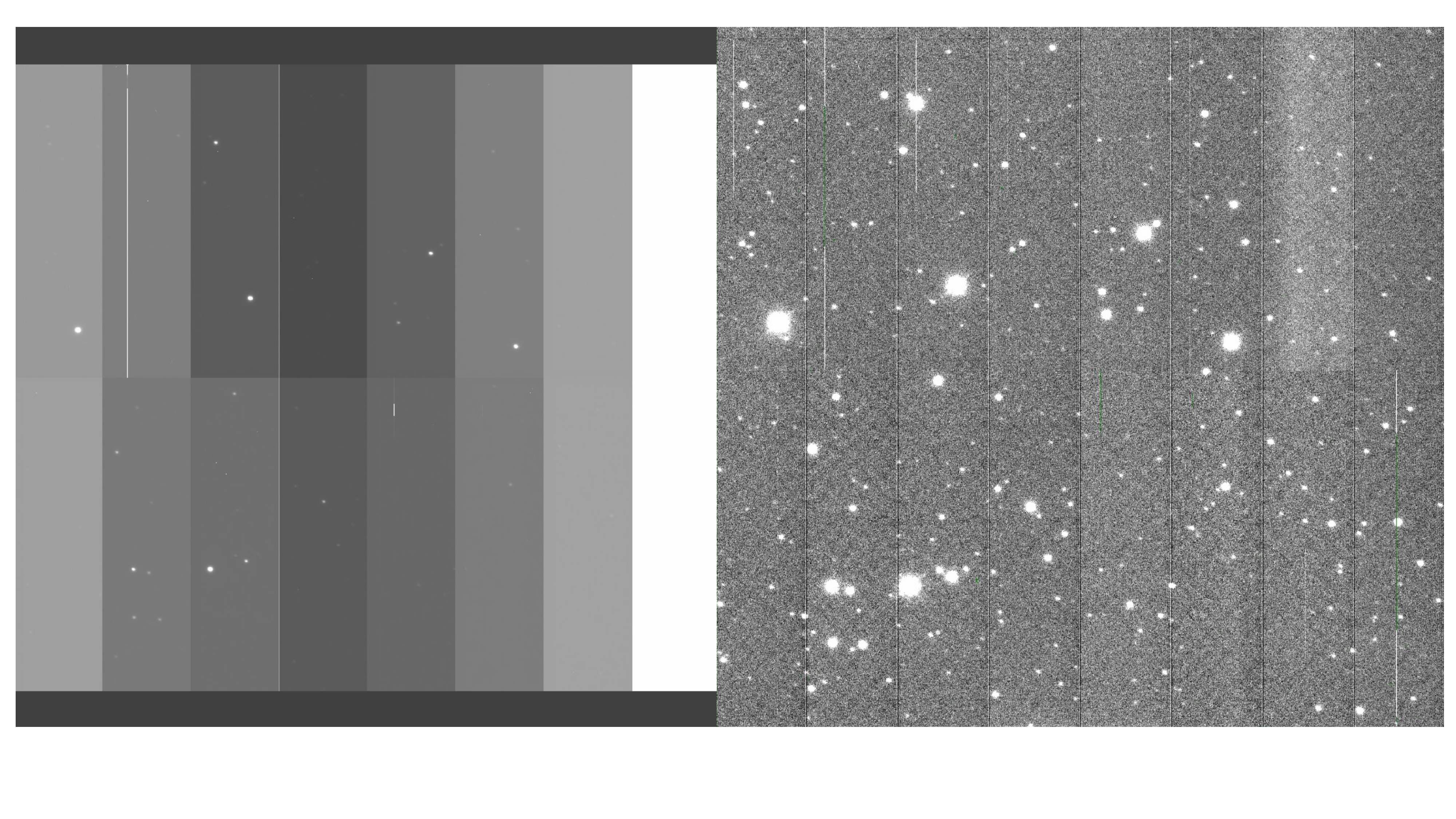}
      \caption{Example of an image before and after partial Instrument Signature Removal (bias, dark, and flat applied).}
      \label{fig:isr_image}
    \end{figure}

\section{Instrument Signature Removal for the LSST}
\label{sec:isr}

\begin{figure}
      \centering
        \includegraphics[width=\linewidth]{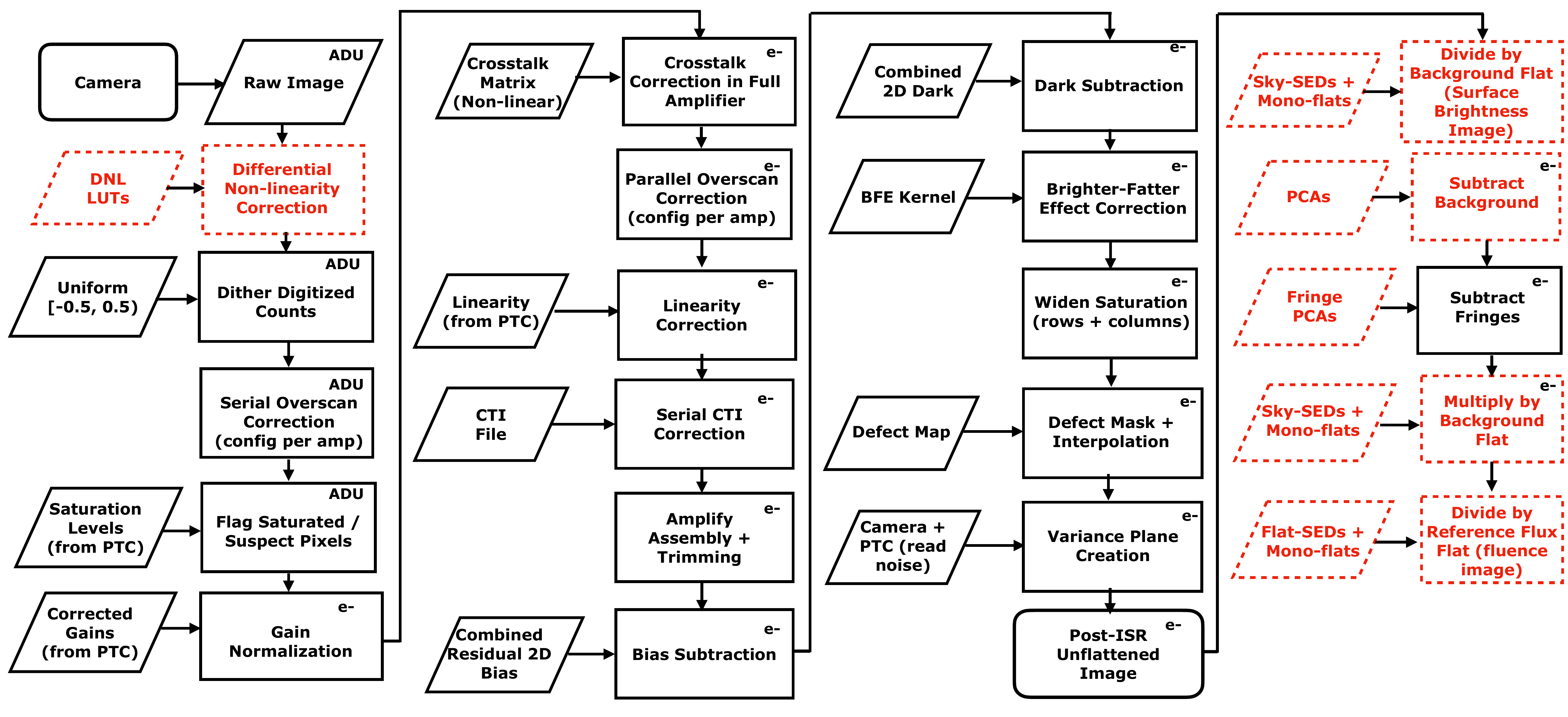}
      \caption{ISR steps and necessary calibration products. Units are displayed in each box, and the steps in red and enclosed in dashed lines are in the process of being implemented {\textcolor{black}{before the start of the LSST, with the exception of synthetic spectro-matched flats, which are not planned for commissioning or Y1 due to their complexity; instead, we will use broadband flats and star-flat-based illumination corrections, similar to Bernstein+17\cite{bernstein17_detendring}.\cite{leanne_p_guy_2024_11110648}}} Credit: Image based on a figure by E. Rykoff. Flowchart steps based on discussions at Rubin Calibration Workshop at SLAC, October 2023.}
       \label{fig:isr_boxes}
\end{figure}

Figure \ref{fig:isr_image} shows an example of an image before and after partial ISR processing, with bias, dark, and flat-field corrections applied. The full sequence of ISR steps, which transform a raw image in {Analog-to-Digital Units} (ADU) into a post-ISR image in e$^{-}$, is displayed in Figure \ref{fig:isr_boxes} and described in the following subsections. These include steps not yet implemented in the pipelines (marked in red and dashed in Figure \ref{fig:isr_boxes}, and indicated with an asterisk in the subsection title). {\textcolor{black}{These steps are expected to be completed by the start of LSST operations, which is scheduled to begin 16 weeks (plus contingency time) after the System First Light milestone, currently planned for July 2025\cite{leanne_p_guy_2024_11110648}. The only exception is synthetic spectro-matched flats, which are not planned for commissioning or Year 1 due to their complexity. Instead, we will use broadband flats and star-flat-based illumination corrections, similar to the approach in Bernstein et al. (2017)\cite{bernstein17_detendring}. The performance of the pipelines will be assessed during commissioning (using data from the Auxiliary Telescope, ComCam, and LSSTCam), implementing state-of-the-art correction algorithms developed within the Commissioning Science units and LSST Science Collaborations, and using the verification metrics mentioned in Sections 3 and 4 for calibration generation and verification.}}

The diagram in Figure \ref{fig:isr_boxes} represents the status of the current ISR algorithms as of version {\tt{w\_2025\_02}} of the LSST Science Pipelines. The ISR code can be found at \url{https://github.com/lsst/ip_isr} 

\subsection{Input validation}
The ISR task receives as input \emph{raw}, unassembled images in Analog-to-Digital (ADU) units. The task then validates that it has the necessary input calibration files to perform the requested ISR steps in the task configuration file.

\subsection{Differential non-linearity correction*}

Differential non-linearity (DNL) in the CCD analog-to-digital converter (ADC) occurs when there is a bias toward certain digital output values (e.g., 0 or 1). Ideally, no ADC levels should be "narrow" {\textcolor{black}{(representing a smaller-than-expected range of input signals) or "wide" (representing a larger-than-expected range)}}, as this would lead to uneven digitization and quantization bias. {\textcolor{black}{While DNL has been measured to be small, it can be monitored through the photon transfer curve (PTC) measurements.}} To address DNL, correction algorithms can be incorporated into the ISR pipelines to generate lookup tables for each potential signal level in ADU.

\subsection{Dither Digitized Counts}
{\textcolor{black}{The ``Dither Digitized Counts" step mitigates the quantization effects introduced during the analog-to-digital conversion (ADC) process. ADC converts continuous voltage signals into discrete integer values, inherently introducing a rounding or truncation bias. This digitization process can lead to systematic errors in the resulting data, particularly when precision is critical.}} 
{\textcolor{black}{To address this, a dithering mode is employed as a principled approach to approximate the original continuous values. After the ADC outputs an integer value, a small random offset, uniformly sampled from the range \([-0.5, 0.5]\), is added to the digitized value. This process effectively reduces the impact of digitization bias by spreading the values over a small range, simulating the original continuous distribution. While this method introduces a minor amount of noise (approximately \(1/12\) ADU on average), it is negligible compared to other noise sources in the system and significantly enhances the fidelity of the data by minimizing quantization artifacts.}}

\subsection{Correction of the serial overscan}
The ISR task performs a serial overscan correction ({the overscan being extra readouts of the serial register after a row has been read out}) to remove amplifier bias levels, which may be unstable with time. For LSSTCam, the routine is {set to define the overscan correction as} the population median per row ({\tt{MEDIAN\_PER\_ROW}}), but other correction options are available ({e.g., via the mean of the overscan region or spline fits}). The first three columns of the overscan are skipped to mitigate the impact of deferred charges (the number of columns to skip is configurable). 

As with other ISR steps, future versions of the code will allow this process to be configured per detector and amplifier.

\subsection{Bad amplifier and \texttt{SATURATED}/\texttt{SUSPECT} pixel masking}

After serial overscan correction, pixels flagged as {\tt{SATURATED}} and {\tt{SUSPECT}} are masked. The appropriate definition of saturation is still under consideration. Potential definitions include the mean pixel full-well, the Photon-Transfer Curve (PTC\cite{Janesick2001}) turnoff, the level at which serial or parallel Charge Transfer Inefficiency (CTI) exceeds requirements, the maximum observed level in a given pixel, and the level at which detectors exhibit significant persistence\cite{sitcomtn086,broughton23}.

\subsection{Gain scaling}
\label{sec:gain}
The process of converting the units of each amplifier from ADU to $e^-$ using the gain measurement involves two steps\cite{sitcomtn086}. Initially, the image is corrected to account for the slight dependence of the gain {on the temperature of the readout electronics board (REB) controlling it} (approximately 0.06\%/°C). This correction is necessary to accommodate any minor temperature shifts and ensure all images can use equivalent calibrations. In the second step, the gains ($e^-$/ADU) are used to convert each amplifier from ADU units to $e^-$ units. After the temperature shifts have been corrected, the PTC analysis is used to measure the gain from each amplifier. The current PTC pipeline in the LSST Science Pipelines takes as input a series of raw flat pairs at different fluxes and has 3 main steps: ``ISR", where instrument-signature removal is applied;  ``extract", where the covariances from flats up to a certain lag (with a default value of 8) using the Fast Fourier Transform algorithm from Appendix A in Astier et al. (2019)\cite{Astier_2019} are measured; and ``Solve", where the measured covariances are fitted to models described in Astier et al. (2019)\cite{Astier_2019} (the models in Equations 16 and 20 of that paper).

\subsection{Correction of the crosstalk from the image to the parallel overscan region and parallel overscan correction}

For certain amplifiers, parallel overscan subtraction is necessary. However, columns affected by hot pixels and saturated sources may bleed into the parallel overscan region, requiring them to be masked and interpolated along the columns. Additionally, these high signals induce electronic crosstalk (see section \ref{ref:crosstalk} below) into overscan regions of other amplifiers, necessitating crosstalk correction of the overscan region. Following these corrections, the ISR task performs the column-by-column parallel overscan correction.

\subsection{Correction of the crosstalk between amplifiers}
\label{ref:crosstalk}

Crosstalk in LSSTCam detectors refers to the phenomenon where a signal from one amplifier ``leaks" into another. This is a significant concern due to the highly parallelized readout of LSST's focal plane, which is designed to read out 3.2 Gpixels in 2 seconds, resulting in a total of 3024 synchronously read-out video channels (16 channels or amplifiers per detector).  The unique combination of high speed, high-resistivity silicon, low power, and close channel spacing makes the LSST readout more prone to electronic crosstalk than previous mosaic cameras. The primary source of crosstalk is capacitive coupling between single-ended video outputs from the CCD, which can occur both within a single CCD and between CCDs in the focal plane. This interference can also originate from the readout electronics themselves. \cite{oconnor15}. 
To correct this, a comprehensive matrix for each detector is required, detailing the effect of each source amplifier on every target amplifier. It's anticipated that crosstalk between CCDs will be less than intra-CCD crosstalk, with inter-raft crosstalk being negligible. Interestingly, the  LSSTCam also shows a unique form of crosstalk between CCD amplifier segments that does not linearly scale with intensity, differing from what would be expected from capacitive coupling alone \cite{snyder21}.

\subsection{Linearity correction}

Signal-chain non-linearity in the LSSTCam detectors\cite{lage19} is corrected per amplifier, and the correction is derived from a dataset used to calculate the PTC:  a set of flat-field images (flat pairs) at different flux levels, monitored by a photodiode. All non-linearity corrections are defined in terms of an additive correction, such that:
\begin{equation*}
    \textrm{corrected value} = \textrm{uncorrected value} + f(\text{uncorrected value})
\end{equation*}
There are currently multiple options for the correction function $f$, including a lookup table, a polynomial function in which a polynomial is used to approximate the inverse of the response function\cite{freudenburg20}, and a spline fit. The current configuration file for LSSTCam employs a multi-node spline (as proposed by P. Astier [LPNH, IN2P3/CNRS]). 

\subsection{Charge-Transfer Inefficiency (CTI) correction}

CTI {is the phenomenon} where the voltage changes that serve to shift the photoelectrons toward the readout amplifiers do so incompletely, resulting in charge that becomes trapped in the silicon.  When this charge is released later, it manifests visually as spurious trails behind every source object. This trailing caused by CTI is particularly problematic as the amount of flux trailed has a nonlinear relationship with the 
{flux and size of the source and the recent illumination history of the detector}. CTI differs between the serial and {parallel transfer} readout directions due to their distinct charge transfer methods and {spacings of wells}. \cite{massey14}
The LSST Science Pipelines implement a detector and segment-dependent CTI correction algorithm described by Snyder et al. (2020)\cite{snyder20} to correct for serial CTI in ITL devices. This algorithm aims to accurately model and account for the position- and flux-dependent nature of CTI trailing seen in LSST images.

\subsection{Amplifier assembly and trimming}

The overscan regions of the images are trimmed, and the 16 amplifiers per detector are assembled into a single CCD image. 

\subsection{Bias subtraction, using a combined bias frame}
\label{sec:bias}
A 2D combined bias frame (commonly referred to in the LSST Science Pipelines as simply a ``bias" frame) must be subtracted during ISR. This frame represents the signal recorded by the CCD when no light is present {and with no integration time}, including the inherent electronic noise of the CCD and the readout system, as well as any other systematic effects present even when no light is hitting the detector. To construct a combined bias frame, a series of individual biases (typically $\approx$ 20 to minimize the noise in the combined bias frame\cite{dmtn101}) or exposures at zero integration time are taken as input.  The overscan, crosstalk, and assembly steps previously discussed are applied to the input zero-second exposure time images, which are then combined using a clipped per-pixel mean ({i.e., an iteratively sigma-clipped mean is computed on the set of individual bias images}) to construct the output bias correction. Crosstalk correction must be performed before bias frame creation to ensure that the crosstalk signal from hot columns is corrected. Otherwise, these signals will imprint on all of the amplifier segments in subsequent detrending steps. 

\subsection{Dark subtraction, using a combined dark frame}

A combined dark image, to be subtracted from science images, is constructed by taking a clipped per-pixel mean from a set of individual dark images, similar to the bias image described in section \ref{sec:bias}. Individual dark images of the entire LSSTCam focal plane are captured with the camera shutter closed, including contributions from dark current and light leakages. Dark images are also used to create defect maps and are taken at different integration times to study dynamic defects or bad columns.

\subsection{Brighter-fatter effect correction}

The brighter-fatter effect (BFE) refers to the observed growth in the sizes of stars with signal level, resulting from the deflection of in-falling charges by transverse electric fields produced by already accumulated charges in the potential wells of pixels.\cite{antilogus14,guyonnet15,gruen2015,broughton23} The BFE leads to correlations between pixels and is calibrated from measured pixel correlations in flat fields, as it smooths out Poisson noise fluctuations at high signal levels. The strength of the BFE can differ along the channel stops or barrier clock gates in a sensor, it can vary mildly with color, and its impact on pixel correlations does not necessarily scale linearly with signal level.

The effect is calibrated from the empirical correlations in the limit of zero signal, at a user-specified signal level, or an average of correlations across a range of signal levels. The pixel correlations are derived from the flat-field image covariances measured out to a lag of 8 pixels, as mentioned in section \ref{sec:gain}. Fully capturing the range of the effect is important for the correction to conserve charge. To verify this, the sum of the pixel correlations out to this distance is checked for consistency with zero. A PTC with at least 1000 points from near zero signal up to saturation is necessary to constrain the anisotropy, magnitude, and range of the effect, as well as the evolution of the strength of the BFE with signal level. When the measurement of correlations at large distances is limited by statistical fluctuations, an empirically measured power law is used to project the pixel correlations out to larger distances. 

The BFE is corrected under the assumption that the derived pixel correlations are proportional to the Laplacian of a constant, unitless, 2-dimensional, scalar kernel, which is assumed to represent the displacement field due to a single accumulated charge \cite{Coulton2018,lage19}. {\textcolor{black}{This kernel is convolved with the measured image to recover a template representing the net gain/loss of charge per pixel (the net sum across the template is forced to be zero); we then use this template to recover the true image.}} The kernel is applied recursively until the total number of shifted charges across the image at each {iteration} falls below a threshold of 1000 electrons. If the kernel accurately models the effect, {no more than 2--3 iterations should be required} to converge to a solution. 

The BFE can empirically vary from amplifier to amplifier or sensor to sensor and mildly from wavelength to wavelength. Therefore, calibrations are derived for each sensor, amplifier, and filter band. However, to avoid discontinuities along amplifier boundaries, the 16 kernels for each channel of a given sensor are averaged to create a single detector-level kernel that is ultimately used to remove the effect from images taken by that sensor. Ultimately, we apply one calibrated kernel per sensor per filter band.

The BFE is small, and the calibration of the BFE can be sensitive to gain estimation as well as the calibration of other artifacts that affect the measured pixel correlations, such as charge transfer inefficiency (CTI), {signal-chain} non-linearity, overscan subtraction, as well as other sensor-specific calibrations. Since the BFE is one of the first instrument signatures to impact an exposure, the correction must be applied near the end of the ISR pipeline. Extreme care must therefore be taken to remove these other effects and validate these other ISR steps before deriving calibrations for the BFE from flat fields and applying the correction to an image. 

{The kernel is initially validated by backward-correcting flat images to linearize the PTC, as a good-quality kernel should be able to recover the Poisson noise in the flat field images. In general, the sub-pixel electric fields inside a flat image do not approximate the sub-pixel electric fields inside a typical science image, and the slope of the second moment vs. magnitude of stellar sources is also checked to be consistent with zero.\cite{astier23,broughton23,lage19}}

\subsection{Defect mask interpolation}

Defects are determined as the outliers of the pixel {value} distributions from dark and flat-field images and are set to the {\tt{BAD}} flag. Individual or combined darks and flats can be used.  The thresholds in darks and flats can be specified in terms of values or standard deviations from the mean. 

\subsection{Variance calculation (variance image construction)}

A weight map is calculated from assembled and trimmed images based on the measured variance per pixel. This is constructed by first dividing each {image segment by the gain ($e^-$/ADU) of that segment} to derive the Poisson variance, and then adding the square of the read noise to derive a map of the empirical uncertainty across an image (the ``Variance Plane Image"). The read noise can be measured empirically from the overscan region of the detector and can also be calculated as a parameter from the fit to the PTC. 

\subsection{Flat-field corrections to convert image to units of fluence*}

After the previous correction steps have been applied, an unflattened, gain-scaled {\tt{post-ISR}} image in e$^{-}$ with instrument signatures removed is obtained. Flat fielding is incorporated into the subsequent photometric calibration process, which aims to estimate the surface brightnesses of celestial sources uniformly across the sky {as would be measured} at the top of the Earth’s atmosphere.\cite{bernstein17_detendring,bernstein18_photometry,sitcomtn086} This process involves constructing a ``background flat" per filter for {image} background subtraction using monochromatic dome flats and sky data ({including background template estimation via, \emph{e.g.,} Principal Component Analysis\cite{bernstein17_detendring})}, along with a ``reference flux flat" per filter and as described in Bernstein et al. (2017)\cite{bernstein17_detendring}. The reference flux flat measures the response to focused light and differs from a dome flat in several respects, including accounting for pixel size variations by removing the effects of transverse electric fields (e.g., tree rings\cite{plazas14b,park17,esteves23} and edge distortions or ``picture frames"\cite{plazas14a,plazas13}), and accounting for differences in { focused and scattered light patterns from flat-field screens and sky sources}, and contamination by scattered light\cite{sitcomtn086}. 

After this flat-fielding process (including fringes that arise from interference effects as light passes through and reflects within the layers of the CCD and its coatings/filters) correction for redder bands\cite{guo23}), a fluence image in e$^{-}$ (following the convention in Bernstein et al. (2017)\cite{bernstein17_detendring}) is obtained, as depicted in the {lower right steps of the flow chart} in Figure \ref{fig:isr_boxes}. The subsequent process of turning this fluence image into a Processed Visit Image in physical units (nJy) involves the construction of a point-spread function model, morphological cosmic-ray finding and masking, and the construction of astrometric\cite{bernstein17_astrometry} and photometric solutions\cite{bernstein18_photometry} (see, e.g., Bosch et al. 2018\cite{bosch2018} for a description of this process in the context of the Hyper Supreme-Cam survey using a previous version of the LSST Science Pipelines).
    \begin{figure}
      \centering
      \includegraphics[width=1.0\linewidth]{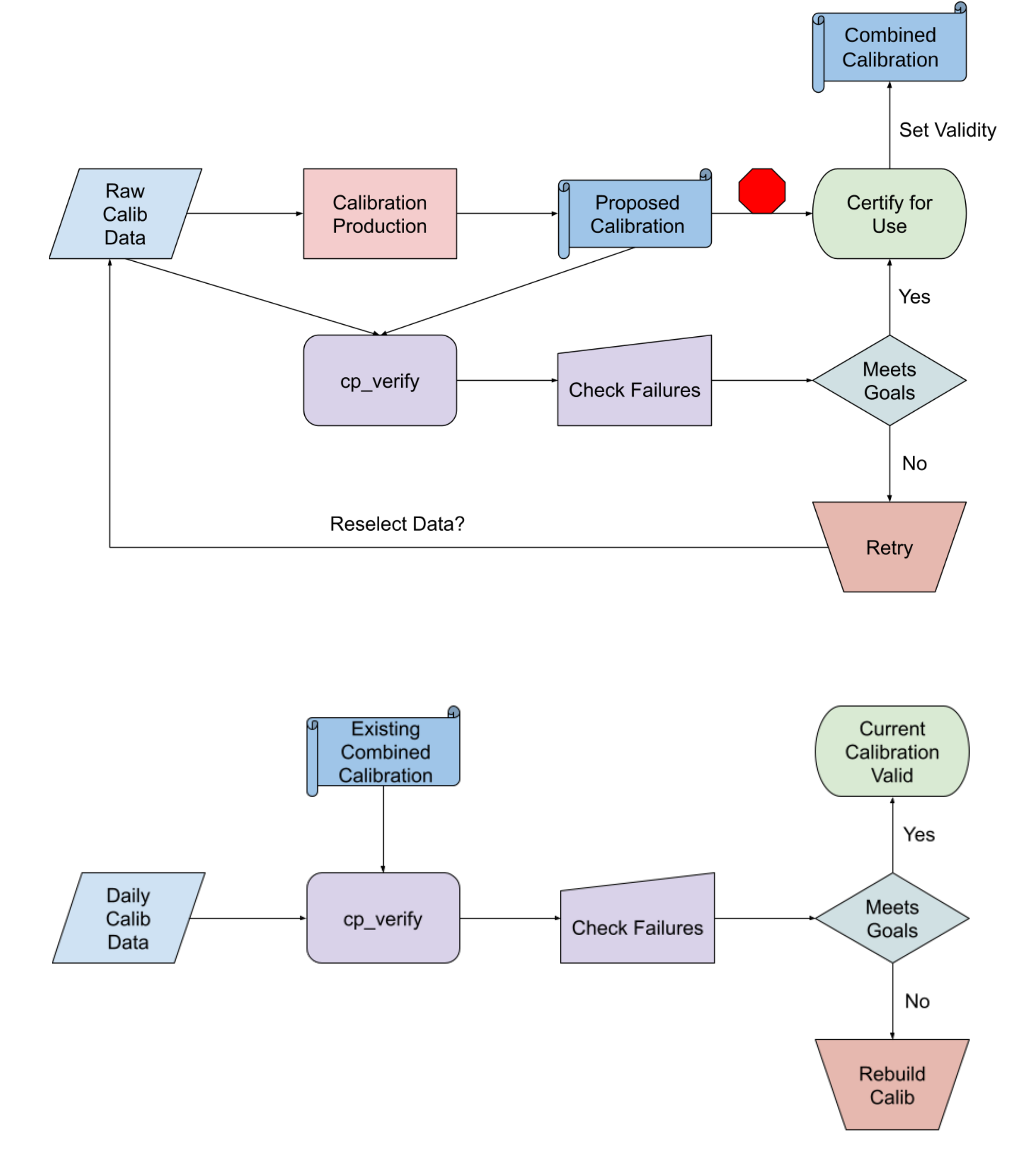}
      \caption{\textbf{Combined and daily} calibration flow charts. Credit: Chris Waters, Data Management Technical Note 222\cite{dmtn222}}
      \label{calib:chart}
    \end{figure}

\section{Calibration Products Production for ISR}
\label{sec:calib}
Calibration products are required to carry out the ISR steps. To validate a new combined calibration for use, metrics are measured on a {residual image that results after applying a given calibration (e.g., a bias-subtracted individual bias)} and compared to expectations and limits (using a series of tests and metrics defined in the Data Management Technical Note 101\cite{dmtn101}, see Table \ref{tab:tests}). This is accomplished via the {\tt{cp\_verify}} package (\url{https://github.com/lsst/cp_verify}). If all metrics are within limits, the calibration is certified for a date range during which it will be used, though the end date may be unknown. 
\begin{table}[H]
\centering
\begin{tabular}{|>{\centering\arraybackslash}m{4cm}|>{\centering\arraybackslash}m{10cm}|}
\hline
\textbf{Calibration Products} & \textbf{Verification Criteria} \\
\hline
\textbf{Bias and Dark} & 
\begin{itemize}
  \item Mean consistent with zero.
  \item Clipped standard deviation consistent with read noise.
  \item Cosmic rays rejected standard deviation consistent with read noise.
\end{itemize} \\
\hline
\textbf{Brighter-fatter correction} & 
\begin{itemize}
  \item Slope of source second-moment size as a function of source magnitude small ($\lessapprox$ 1\%).
\end{itemize} \\
\hline
\end{tabular}
\caption{Examples of {\tt{cp\_verify}} tests for calibration verification. Bias: using bias-corrected individual bias exposures. Dark: using bias- and dark-corrected individual dark exposures. Brighter-fatter correction: using full ISR-processed science exposures.}
\label{tab:tests}
\end{table}
These validations are also used with daily calibrations to monitor the stability of the camera and telescope by confirming that existing combined calibrations remain suitable.  Figure \ref{calib:chart} shows the relationship between construction, validation, and usage of combined and \textbf{\textcolor{black}{daily}} calibrations. The steps of the process can be summarized as follows (see ``Data Management Technical Note 222: Calibration Generation, Verification, Acceptance, and Certification"\cite{dmtn222}):

\begin{enumerate}
    \item \textbf{Generation}: {\textcolor{black}{Calibration products are generated using raw exposures through the {\tt{cp\_pipe}} package, ensuring precursor dependencies (e.g., photon transfer curves for gain, brighter-fatter kernels, etc.) are taken into account. The input exposures are selected based on calibration type (e.g., bias, dark, flat) and aligned with the required observation types or sequences.}}

    \item \textbf{Verification}: {\textcolor{black}{The proposed calibration is evaluated using the {\tt{cp\_verify}} package, which measures quality metrics as defined in the Data Management Technical Note 101\cite{dmtn101}. These metrics include diagnostic plots and extended tests to validate the calibration's performance across different input exposures. Metrics are compared to predefined limits, ensuring compatibility with science requirements and detecting potential anomalies.}}

    \item \textbf{Certification}: {\textcolor{black}{Once verified, calibrations are certified for a specific validity range. This typically includes a starting date that corresponds to system changes, while the end date remains open-ended unless a new calibration supersedes it. Certification is performed via database operations that link the calibration to the respective datasets in the repository.}}

    \item \textbf{Approval}: {\textcolor{black}{The Telescope and Auxiliary Instrumentation Calibration Acceptance Board (TAXICAB) reviews verification reports and approves the calibration based on consensus. If required, additional processing steps are defined before approval is granted.}}

    \item \textbf{Distribution}: {\textcolor{black}{Approved calibrations are distributed to all relevant repositories using the Butler framework (see next section), ensuring consistent usage across the system.}} 
\end{enumerate}

    \begin{figure}
      \centering
         \begin{minipage}{\textwidth}
        \centering
        \includegraphics[width=1.0\linewidth]{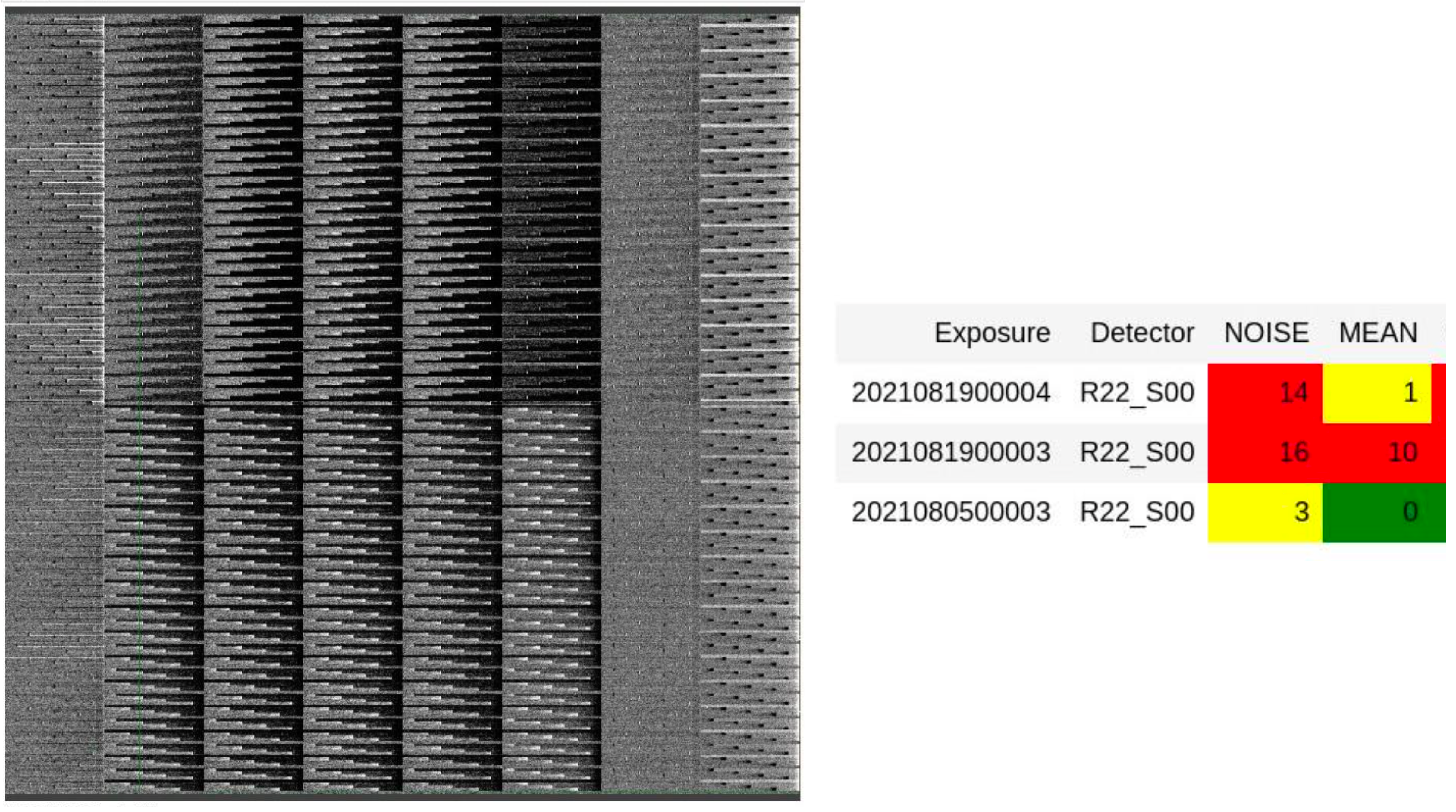}
    \end{minipage}
      \caption{{\tt{cp\_verify}} identifying a defective combined bias from a particular detector due to an acquisition problem, as indicated in the ``failure table" in the right, using the tests of Table \ref{tab:tests}: {\tt{NOISE}} corresponds to ``Clipped standard deviation consistent with read noise" and {{\tt{MEAN}} corresponds to ``Mean consistent with zero".}}
      \label{cp_verify_example}
    \end{figure}
Figure \ref{cp_verify_example} shows an example of a defective bias failing the tests defined in {\tt{cp\_verify}}.

{\textcolor{black}{The frequency of updates to calibration products depends on the type and purpose of the calibration. Daily calibrations, including bias, dark, and flat frames, are acquired to monitor the stability of the camera and telescope systems but are not automatically used to generate new combined calibrations. Periodic calibrations, such as dense PTCs and the brighter-fatter kernel, are constructed less frequently (on a 1-2 month cadence) due to their reliance on dense observational sequences of flats or specific measurement conditions, though they are validated periodically via {\tt{cp\_verify}} to ensure continued accuracy. Calibrations that are not calculated from available data (like quantum efficiency curves) or are relatively static (like manual defect masks) are called ``curated calibrations." These calibrations are updated only as needed, often in response to hardware or software changes, and require specialized procedures and equipment. Additionally, calibration frames linked to known system changes---such as upgrades to the camera or telescope power cycles, mirror coating, etc---are updated with a defined starting validity date, while older calibrations remain in use until the new ones are validated. For more details, see the Rubin ``Data Management Technical Note 222: Calibration Generation, Verification, Acceptance, and Certification"\cite{dmtn222}}}.

\section{The LSST Data Butler, Calibration Collections, and Calibration Pipelines}
\label{sec:butler}

\begin{figure}
      \centering
         \begin{minipage}{\textwidth}
        \centering
        \includegraphics[width=1.0\linewidth]{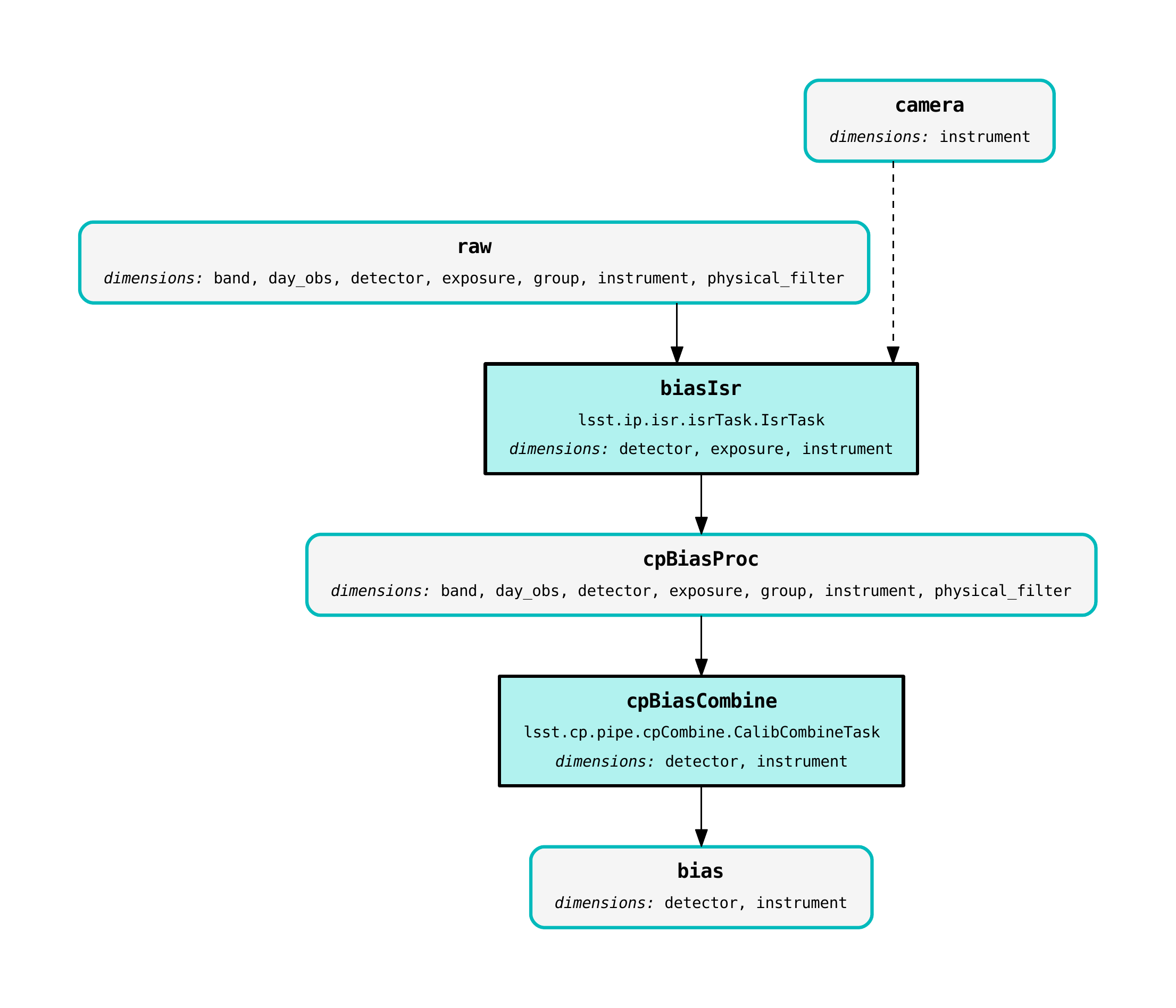}
    \end{minipage}
      \caption{{\tt{PipelineTasks}} in the bias pipeline to produce a combined bias. Diagram credit: L. Kelvin and the LSST Science Pipelines.}
      \label{fig:bias_pipeline}
    \end{figure}

The Rubin data management system uses a ``data Butler" to organize and locate datasets\cite{jenness22}. This Butler groups data into ``collections," with a special ``calibration collection" that associates calibration datasets with a validity time range.  When image processing is started, the Butler uses these validity time ranges to match the supplied calibration data to the images to be processed. Curated calibrations are stored in text files in GitHub repositories with a directory structure that encodes their validity ranges. For example, \url{https://github.com/lsst/obs_lsst_data/tree/main/lsstCam} stores transmission curves for LSSTCam. \\
Processing is done via {\tt{PipelineTasks}}. Each {\tt{PipelineTask}} provides one processing step, {and several can be combined} into a full pipeline defined in a YAML file.
Figure \ref{fig:bias_pipeline} shows a diagram depicting the {\tt{PipelineTasks}} involved in the creation of a combined bias using the bias pipeline. {The \emph{dimensions} are the key variables or parameters that define the characteristics of a Butler dataset, such as its observation time, instrument configuration, target coordinates, filter, etc., that allow it to be uniquely identified and located within the data archive system.\cite{jenness22}} The ISR {\tt{PipelineTasks}} for LSSTCam provide the functionality to remove systematic errors from the raw instrument data, using the calibration products discussed above. Multiple ISR {\tt{PipelineTasks}} with different configurations can be defined to handle different instruments.

\section{Conclusion}

{\textcolor{black}{The Rubin Legacy Survey of Space and Time (LSST) represents an unprecedented opportunity to revolutionize our understanding of the universe through precise measurements of astrophysical phenomena. Achieving the ambitious scientific goals of the LSST requires meticulous control of systematic errors, starting with the Instrument Signature Removal (ISR) pipelines. This paper has described the current state of the ISR algorithms in the LSST Science Pipelines, highlighting their role in transforming raw CCD data into scientifically usable images by correcting for effects such as bias, dark current, crosstalk, and the brighter-fatter effect.}}

{\textcolor{black}{After calibration products for ISR are generated, they are verified against predefined metrics using the \texttt{cp\_verify} package and certified and accepted for use within specified validity ranges. This process ensures that calibrations meet scientific requirements while allowing for periodic updates to account for long-term trends and hardware changes. Daily calibrations are employed to monitor the stability of the system.}}

{\textcolor{black}{The ISR pipelines, along with the calibration production and verification processes, continue to be refined in preparation for the start of LSST operations. All remaining ISR steps and calibration processes, including those currently under development, are expected to be completed by the start of LSST operations (16 weeks—plus contingency time—after the scheduled System First Light milestone, currently planned for July 2025\cite{leanne_p_guy_2024_11110648}). These efforts will ensure that the LSST Science Pipelines are fully equipped to meet the survey's stringent performance requirements, enabling transformative science for the next decade and beyond.}}

\subsection*{Disclosures}
There are no conflicts of interest. 

\subsection* {Code, Data, and Materials Availability}
The LSST Science Pipelines (\url{pipelines.lsst.io}) code described in this article is open-source and available in the following repositories: \url{https://github.com/lsst/ip_isr}, \url{https://github.com/lsst/cp_pipe}, and \url{https://github.com/lsst/cp_verify}. 

\subsection* {Acknowledgments}
The work of AAPM was supported by the U.S. Department of Energy under contract number DE-AC02-76SF00515.
This paper makes use of LSST Science Pipelines software developed by the Vera C. Rubin Observatory. We thank the Rubin Observatory for making their code available as free software at {\url{https://pipelines.lsst.io}}. {This material or work is supported in part by the National Science Foundation through Cooperative Agreement AST-1258333 and Cooperative Support Agreement AST1836783 managed by the Association of Universities for Research in Astronomy (AURA), and the Department of Energy under Contract No. DE-AC02-76SF00515 with the SLAC National Accelerator Laboratory managed by Stanford University.} AAPM  thanks the Department of Physics of Harvard University and the Laboratory of Particle Astrophysics and Cosmology, the Cosmology Group at Boston University, and the Department of Physics at Washington University in St. Louis for their hospitality during the preparation of this paper. {We thank S. Digel (SLAC National Accelerator Laboratory) for serving as Rubin's internal reviewer and providing comments that improved the manuscript. We thank T. Jenness (Rubin), {\v{Z.} Ivezi{\'c}} (Rubin, University of Washington) {\textcolor{black}{and the anonymous JATIS referees}} for providing comments that improved the manuscript. {\textcolor{black}{During the preparation of this work, the first author used ChatGPT-4o to overcome language barriers posed by English as a second language \cite{Giglio2023} by improving English grammar and writing. After using this tool/service, the authors reviewed and edited the content as needed and take full responsibility for the final publication.}}


\bibliography{article}   

\begin{thebibliography}{10}

\bibitem{Ivezi__2019}
{\v{Z}}.~{Ivezi{\'c}}, S.~M. {Kahn}, J.~A. {Tyson}, {\em et~al.}, ``{LSST: From Science Drivers to Reference Design and Anticipated Data Products},'' {\em The Astrophysical Journal} {\bf 873}, 111  (2019).

\bibitem{holland03}
S.~E. {Holland}, D.~E. {Groom}, N.~P. {Palaio}, {\em et~al.}, ``{Fully depleted, back-illuminated charge-coupled devices fabricated on high-resistivity silicon},'' {\em IEEE Transactions on Electron Devices} {\bf 50}, 225--238  (2003).

\bibitem{holland09}
S.~E. {Holland}, W.~F. {Kolbe}, and C.~J. {Bebek}, ``{Device Design for a 12.3-Megapixel, Fully Depleted, Back-Illuminated, High-Voltage Compatible Charge-Coupled Device},'' {\em IEEE Transactions on Electron Devices} {\bf 56}, 2612--2622  (2009).

\bibitem{holland14}
S.~E. {Holland}, C.~J. {Bebek}, W.~F. {Kolbe}, {\em et~al.}, ``{Physics of fully depleted CCDs},'' {\em Journal of Instrumentation} {\bf 9}, C03057  (2014).

\bibitem{O'Connor16}
P.~{O'Connor}, P.~{Antilogus}, P.~{Doherty}, {\em et~al.}, ``{Integrated system tests of the LSST raft tower modules},'' in {\em High Energy, Optical, and Infrared Detectors for Astronomy VII},  A.~D. {Holland} and J.~{Beletic}, Eds., {\em Society of Photo-Optical Instrumentation Engineers (SPIE) Conference Series} {\bf 9915}, 99150X  (2016).

\bibitem{Arndt10}
K.~{Arndt}, V.~{Riot}, E.~{Alagoz}, {\em et~al.}, ``{The LSST camera corner raft conceptual design: a front-end for guiding and wavefront sensing},'' in {\em Adaptive Optics Systems II},  B.~L. {Ellerbroek}, M.~{Hart}, N.~{Hubin}, {\em et~al.}, Eds., {\em Society of Photo-Optical Instrumentation Engineers (SPIE) Conference Series} {\bf 7736}, 773662  (2010).

\bibitem{lesser17}
M.~{Lesser} and D.~{Ouellette}, ``{Results from STA/ITL fully depleted CCDs for LSST},'' {\em Journal of Instrumentation} {\bf 12}, C03080  (2017).

\bibitem{itl_website}
ITL, ``\url{https://www.itl.arizona.edu/},''  (2025).
\newblock Accessed: 29-Jan-2025.

\bibitem{e2v_website}
e2v, ``\url{https://www.teledyneimaging.com/en/aerospace-and-defense/products/sensors-overview/ccd/ccd250-82/},''  (2025).
\newblock Accessed: 29-Jan-2025.

\bibitem{lsst2011}
{\v{Z}}.~{Ivezi{\'c}} and {The LSST Science Collaboration}., ``{The LSST System Science Requirements Document},'' {\em The LSST System Science Requirements Document LPM-17}   (2011).

\bibitem{desc2018}
{The LSST Dark Energy Science Collaboration}, R.~{Mandelbaum}, T.~{Eifler}, {\em et~al.}, ``{The LSST Dark Energy Science Collaboration (DESC) Science Requirements Document},'' {\em arXiv e-prints} , arXiv:1809.01669  (2018).

\bibitem{lsstDPDP}
M.~{Juri{\'c}} and {et al.}., ``{Data Products Definition Document},'' {\em The LSST Data Products Definition Document LSE-163}   (2023).

\bibitem{comcam_commissioning}
R.~Observatory, ``Comcam commissioning: \url{https://community.lsst.org/t/2024-12-13-on-sky-commissioning-update/9644},''  (2025).
\newblock Accessed: 29-Jan-2025.

\bibitem{howard2018}
J.~{Howard}, K.~{Reil}, C.~{Claver}, {\em et~al.}, ``{The LSST commissioning camera status and progress},'' in {\em Ground-based and Airborne Telescopes VII},  H.~K. {Marshall} and J.~{Spyromilio}, Eds., {\em Society of Photo-Optical Instrumentation Engineers (SPIE) Conference Series} {\bf 10700}, 107003D  (2018).

\bibitem{schutt2025}
T.~{Schutt}, M.~{Jarvis}, A.~{Roodman}, {\em et~al.}, ``{Dark Energy Survey Year 6 Results: Point-Spread Function Modeling},'' {\em arXiv e-prints} , arXiv:2501.05781  (2025).

\bibitem{li2022}
X.~{Li}, H.~{Miyatake}, W.~{Luo}, {\em et~al.}, ``{The three-year shear catalog of the Subaru Hyper Suprime-Cam SSP Survey},'' {\em Publications of the Astronomical Society of Japan} {\bf 74}, 421--459  (2022).

\bibitem{jarvis2021}
M.~{Jarvis}, G.~M. {Bernstein}, A.~{Amon}, {\em et~al.}, ``{Dark Energy Survey year 3 results: point spread function modelling},'' {\em Monthly Notices of the Royal Astronomical Society} {\bf 501}, 1282--1299  (2021).

\bibitem{bosch2018}
J.~{Bosch}, R.~{Armstrong}, S.~{Bickerton}, {\em et~al.}, ``{The Hyper Suprime-Cam software pipeline},'' {\em Publications of the Astronomical Society of Japan} {\bf 70}, S5  (2018).

\bibitem{bosch2019}
J.~{Bosch}, Y.~{AlSayyad}, R.~{Armstrong}, {\em et~al.}, ``{An Overview of the LSST Image Processing Pipelines},'' in {\em Astronomical Data Analysis Software and Systems XXVII},  P.~J. {Teuben}, M.~W. {Pound}, B.~A. {Thomas}, {\em et~al.}, Eds., {\em Astronomical Society of the Pacific Conference Series} {\bf 523}, 521  (2019).

\bibitem{leanne_p_guy_2024_11110648}
L.~P. Guy, K.~Bechtol, E.~Bellm, {\em et~al.}, {\em RTN-011 - Rubin Observatory Plans for an Early Science Program:\url{{https://doi.org/10.5281/zenodo.11110648}}}  (2024).

\bibitem{bernstein17_detendring}
G.~M. {Bernstein}, T.~M.~C. {Abbott}, S.~{Desai}, {\em et~al.}, ``{Instrumental response model and detrending for the Dark Energy Camera},'' {\em Publications of the Astronomical Society of the Pacific} {\bf 129}, 114502  (2017).

\bibitem{Janesick2001}
J.~R. Janesick, {\em {S}cientific charge-coupled devices}, SPIE, Bellingham  (2001).

\bibitem{sitcomtn086}
{P. Fagrelius} and {E. Rykoff}, ``{Rubin Baseline Calibration Plan},''  (2023).
\newblock {Vera C. Rubin Observatory Commissioning Technical Note SITCOMTN-086}.

\bibitem{broughton23}
A.~{Broughton}, Y.~{Utsumi}, A.~{Plazas Malag{\'o}n}, {\em et~al.}, ``{Mitigation of the Brighter-Fatter Effect in the LSST Camera},'' {\em arXiv e-prints} , arXiv:2312.03115  (2023).

\bibitem{Astier_2019}
P.~{Astier}, P.~{Antilogus}, C.~{Juramy}, {\em et~al.}, ``{The shape of the photon transfer curve of CCD sensors},'' {\em AAP} {\bf 629}, A36  (2019).

\bibitem{oconnor15}
P.~{O'Connor}, ``{Crosstalk in multi-output CCDs for LSST},'' {\em Journal of Instrumentation} {\bf 10}, C05010  (2015).

\bibitem{snyder21}
A.~{Snyder}, A.~{Barrau}, A.~{Bradshaw}, {\em et~al.}, ``{Laboratory Measurements of Instrumental Signatures of the LSST Camera Focal Plane},'' {\em arXiv e-prints} , arXiv:2101.01281  (2021).

\bibitem{lage19}
C.~{Lage}, ``{Linearity and correction of the BF effect in LSST sensors},'' {\em arXiv e-prints} , arXiv:1911.09567  (2019).

\bibitem{freudenburg20}
J.~K.~C. {Freudenburg}, J.~J. {Givans}, A.~{Choi}, {\em et~al.}, ``{Brighter-fatter Effect in Near-infrared Detectors{\textemdash}III. Fourier-domain Treatment of Flat Field Correlations and Application to WFIRST},'' {\em PASP} {\bf 132}, 074504  (2020).

\bibitem{massey14}
R.~{Massey}, T.~{Schrabback}, O.~{Cordes}, {\em et~al.}, ``{An improved model of charge transfer inefficiency and correction algorithm for the Hubble Space Telescope},'' {\em Monthly Notices of the Royal Astronomical Society} {\bf 439}, 887--907  (2014).

\bibitem{snyder20}
A.~{Snyder} and A.~{Roodman}, ``{Investigation of Deferred Charge Effects in LSST ITL Sensors},'' {\em arXiv e-prints} , arXiv:2001.03223  (2020).

\bibitem{dmtn101}
{Lupton, R.} and {Plazas Malag{\'o}n, A.}, ``{Verifying LSST Calibration Data Products},''  (2018).
\newblock {Vera C. Rubin Observatory Data Management Technical Note DMTN-101}.

\bibitem{antilogus14}
P.~Antilogus, P.~Astier, P.~Doherty, {\em et~al.}, ``The brighter-fatter effect and pixel correlations in ccd sensors,'' {\em Journal of Instrumentation} {\bf 9}, C03048–C03048  (2014).

\bibitem{guyonnet15}
A.~Guyonnet, P.~Astier, P.~Antilogus, {\em et~al.}, ``Evidence for self-interaction of charge distribution in charge-coupled devices,'' {\em AAP} {\bf 575}, A41  (2015).

\bibitem{gruen2015}
D.~{Gruen}, G.~M. {Bernstein}, M.~{Jarvis}, {\em et~al.}, ``{Characterization and correction of charge-induced pixel shifts in DECam},'' {\em Journal of Instrumentation} {\bf 10}, C05032  (2015).

\bibitem{Coulton2018}
W.~R. {Coulton}, R.~{Armstrong}, K.~M. {Smith}, {\em et~al.}, ``{Exploring the Brighter-fatter Effect with the Hyper Suprime-Cam},'' {\em Astronomical Journal} {\bf 155}, 258  (2018).

\bibitem{astier23}
P.~{Astier} and N.~{Regnault}, ``{Correction of the brighter-fatter effect on the CCDs of Hyper Suprime-Cam},'' {\em Astronomy and Astrophysics} {\bf 670}, A118  (2023).

\bibitem{bernstein18_photometry}
G.~M. {Bernstein}, T.~M.~C. {Abbott}, R.~{Armstrong}, {\em et~al.}, ``{Photometric Characterization of the Dark Energy Camera},'' {\em Publications of the Astronomical Society of the Pacific} {\bf 130}, 054501  (2018).

\bibitem{plazas14b}
A.~A. {Plazas}, G.~M. {Bernstein}, and E.~S. {Sheldon}, ``{On-Sky Measurements of the Transverse Electric Fields' Effects in the Dark Energy Camera CCDs},'' {\em Publications of the Astronomical Society of the Pacific} {\bf 126}, 750--760  (2014).

\bibitem{park17}
H.~Y. {Park}, A.~{Nomerotski}, and D.~{Tsybychev}, ``{Properties of tree rings in LSST sensors},'' {\em Journal of Instrumentation} {\bf 12}, C05015  (2017).

\bibitem{esteves23}
J.~H. {Esteves}, Y.~{Utsumi}, A.~{Snyder}, {\em et~al.}, ``{Photometry, Centroid and Point-spread Function Measurements in the LSST Camera Focal Plane Using Artificial Stars},'' {\em Publications of the Astronomical Society of the Pacific} {\bf 135}, 115003  (2023).

\bibitem{plazas14a}
A.~A. {Plazas}, G.~M. {Bernstein}, and E.~S. {Sheldon}, ``{Transverse electric fields' effects in the Dark Energy Camera CCDs},'' {\em Journal of Instrumentation} {\bf 9}, C04001  (2014).

\bibitem{plazas13}
A.~A. {Plazas Malag\'on}, ``Transverse electric fields effects in decam devices: tree rings and glowing edges,'' {\em Precision Astronomy with Fully-Depleted CCDs (PACCDs), Brookhaven National Laboratory, November, 2013}, Zenodo, DOI: 10.5281/zenodo.10064122  (2013).

\bibitem{guo23}
Z.~{Guo}, C.~W. {Walter}, C.~{Lage}, {\em et~al.}, ``{Fringing Analysis and Simulation for the Vera C. Rubin Observatory's Legacy Survey of Space and Time},'' {\em Publications of the Astronomical Society of the Pacific} {\bf 135}, 034503  (2023).

\bibitem{bernstein17_astrometry}
G.~M. {Bernstein}, R.~{Armstrong}, A.~A. {Plazas}, {\em et~al.}, ``{Astrometric Calibration and Performance of the Dark Energy Camera},'' {\em Publications of the Astronomical Society of the Pacific} {\bf 129}, 074503  (2017).

\bibitem{dmtn222}
C.~Waters, ``{Calibration Generation, Verification, Acceptance, and Certification.},''  (2023).
\newblock {Vera C. Rubin Observatory Data Management Technical Note DMTN-222}.

\bibitem{jenness22}
T.~{Jenness}, J.~F. {Bosch}, A.~{Salnikov}, {\em et~al.}, ``{The Vera C. Rubin Observatory Data Butler and pipeline execution system},'' in {\em Software and Cyberinfrastructure for Astronomy VII},  {\em Society of Photo-Optical Instrumentation Engineers (SPIE) Conference Series} {\bf 12189}, 1218911  (2022).

\bibitem{Giglio2023}
A.~D. Giglio and M.~U. P.~D. Costa, ``The use of artificial intelligence to improve the scientific writing of non-native english speakers,'' {\em Revista da Associacao Medica Brasileira (1992)} {\bf 69}(9), e20230560  (2023).

\end{thebibliography}
\bibliographystyle{spiejour}   

\listoffigures
\listoftables

\end{document}